# Anisotropic Photostriction and Strain-modulated Carrier Lifetimes in Orthorhombic Semiconductors

Jianxin Yu[1,+], Kun Yang[1,+], Jiawen Li[1,+], Sheng Meng[2,3], Xinghua Shi[1], Jin Zhang[1,*]

[1]*Laboratory of Theoretical and Computational Nanoscience, National Center for Nanoscience and Technology, Chinese Academy of Sciences. Beijing 100190, P. R. China*

[2]*Beijing National Laboratory for Condensed Matter Physics, and Institute of Physics, Chinese Academy of Sciences, Beijing 100190, P. R. China*

[3]*Songshan Lake Materials Laboratory, Dongguan, Guangdong 523808, P. R. China*

\* Corresponding author: Jin Zhang (jinzhang@nanoctr.cn)

We demonstrate anisotropic photostriction in two-dimensional orthorhombic semiconductors using time-dependent density functional theory. By tracing the dynamics of photoexcited carriers, we establish a quantitative link between carrier density and lattice deformation in layered black phosphorus and germanium selenides. The structural response exhibits significant anisotropy, featuring lattice expansion along the armchair direction and contraction along the zigzag direction, which is attributed to the interplay between charge redistribution and intrinsic lattice anisotropy. Both the magnitude and orientation of the photostrictive strains can be tuned by photodoping densities, enabling precise control over the photoinduced response. Notably, the photoinduced strains significantly increase carrier recombination lifetimes by suppressing nonradiative recombination, primarily due to the enlarged bandgap and weakened nonadiabatic coupling. These results provide microscopic insight into the origin of anisotropic photostriction in low-dimensional systems and lay the groundwork for light-controllable, directionally sensitive optomechanical devices at the atomic scale.

# Introduction.

Photostriction, the light-induced, nonthermal expansion or contraction of a crystal lattice, has emerged as a fascinating route for converting optical stimuli into mechanical deformations, holding promise for applications in memories, optical sensors, and energy-harvesting devices[1-3]. In semiconductors, photoexcited carriers couple to the lattice through electron-phonon interactions, inducing collective deformation without significant heating[1]. The photostriction effect primarily originates from the electronic volume effect[4], where photo-induced redistribution of electronic occupancies modifies bonding properties and generates strains. Recent experiments and theoretical research have shown that photostriction is universal across ferroelectrics[5], complex oxides[6, 7], hybrid perovskites[2, 8, 9]. The cooperative interplays between electronic and lattice degrees of freedom give rise to tunable strains, opening avenues toward actively controllable optomechanical and light-responsive systems [10, 11].

Two-dimensional (2D) orthorhombic materials have emerged as fertile platforms for discovering unconventional anisotropic phenomena. Among them, black phosphorus (BP) and germanium selenides (GeSe) are particularly intriguing for orientation-dependent physics, including charge transport, and exciton behaviours[12-14]. It is revealed that the intrinsic anisotropy of monolayer BP permits in-plane manipulation of light polarization, offering a promising route for integration with silicon photonic devices[14]. Furthermore, monochalcogenides have been shown to possess intrinsic ferroelectricity and markedly second-harmonic-generation polarization[5, 15], features that make them attractive candidates for nonlinear optoelectronic technologies. Mechanically, anisotropic tensile superelasticity has been observed under mechanical stimuli in GeSe[16], with analogous responses reported in other chalcogenides sharing similar crystal structures and their doped counterparts[17]. Within this context, photostriction emerges as a powerful yet underexplored route to contactless control of lattice and functionality. However, the light-induced structural dynamics in 2D orthorhombic materials remain largely uncharted, leaving open fundamental questions about how the lattice anisotropy governs their ultrafast structural evolution and functional transformation.

In this work, we employ first-principles simulations based on time-dependent density functional theory (TDDFT) to investigate the directional photostriction phenomena in 2D orthorhombic materials. We systematically analyze the lattice response following carrier excitations. Our calculations reveal that compounds such as BP and GeSe exhibit anisotropic structural deformation upon optical excitation driven by the interplay between charge redistribution and the elastic anisotropy of the lattice. This photostrictive behavior manifests as an expansion along the armchair direction accompanied by a simultaneous contraction along the zigzag axis. The biaxial strain increases the bandgap and shifts the band edge of optical absorption spectra. Furthermore, the strain suppresses nonradiative recombination, significantly prolonging carrier lifetimes. The findings provide a microscopic insight into the anisotropic photostriction effect in 2D systems and suggest a pathway for orientation-dependent optomechanical functionalities at the nanoscale.

**Results and Discussions.**

Figure 1 illustrates the photostriction effect, contrasting isotropic and anisotropic strain responses due to intrinsic symmetries of materials. Upon strong illumination, 2D semiconductors absorb incident photons and generate a transient population of non-equilibrium electron-hole pairs. The photoexcited carriers modulate the local bonding landscape and redistribute internal stress, which in turn drives measurable lattice deformations. In hexagonal-lattice materials (e.g., 2D transition metal dichalcogenides)[11, 18], this carrier-lattice interaction typically results in an isotropic strain. In contrast, orthorhombic-lattice systems (e.g., group-IV monochalcogenide)[19] manifest anisotropic lattice response. Deformations are directionally different contraction or expansion. The inherent directional asymmetry of the lattice causes an uneven distribution of electronic states and deformation potentials. The magnitude of the photostriction is intensely governed by the underlying structures, engendering qualitatively distinct photostrictive effects in isotropic versus anisotropic systems, enabling them as directionally tunable optomechanical applications.

Black phosphorus consists of van der Waals stacked phosphorene layers and crystallizes in an orthorhombic structure. Each layer of BP consists of phosphorus atoms covalently bonded in a folded honeycomb configuration, displaying two distinct directions: the armchair and zigzag axes, as shown in Fig. 2a-b. The bandgap of the bulk phase is 0.15 eV on the PBE level in Fig. 2c, coherently with previous simulations[20]. The band dispersion reveals clear anisotropy, with the armchair direction displaying steeper band dispersion, bringing about a smaller effective mass and higher charge carrier mobility[20].

Analogously, layered GeSe manifests the puckered orthorhombic lattice. Each monolayer comprises zigzag chains of Ge and Se atoms arranged in a puckered configuration (Fig. 2d-e). The optical spectra of pristine GeSe also have an obvious response to polarization of armchair (1.35 eV) or zigzag (1.98 eV) directions. GeSe displays an indirect bandgap of 0.86 eV (Fig. 2f), whereas previous experimental studies have reported values around 1.2 eV.[21]

When exposed to above-bandgap laser pulses, photoexcited carriers redistribute the internal electrostatic potential landscape and renormalize the interatomic forces, leading to lattice distortions. This anisotropic photostriction is not only observed in layered orthorhombic semiconductors but also controllable via light polarization, intensity, and frequency.

**Photoinduced lattice deformation.** To evaluate photocarrier dynamics in layered orthorhombic semiconductors, we conduct real-time time-dependent density functional theory (TDDFT) simulations[22-27]. Laser excitations are modeled using Gaussian-enveloped pulses: $E(t) = E_0 \cos(\omega t) \exp\left[-\frac{(t-t_0)^2}{2\sigma^2}\right]$, where $E_0$, $\omega$, $t_0$, and $\sigma$ denote peak field strength, photon energy, temporal center, and pulse width, respectively. Optical stimulation directly excites electrons, initiating structural reorganization via electron-phonon coupling. Our TDDFT simulation employs 2 eV photons[28-30] at varied intensities, which exceeds the bandgaps of the two materials. Photodoping density quantifies electrons of valence-to-conduction band transitions after pulse termination, computed through ground-state wavefunction projections.

Firstly, we investigate the response of the photodoping carrier density to electric field strengths in BP, as shown in Fig. S1. As for $E_0 = 0.005 \, V/Å$, a relatively small density of approximately 0.04 e/atom is introduced, reflecting the mild perturbation of the electronic states. As the $E_0$ increases to $0.01 \, V/Å$, the photodoping carrier rises to 0.12 e/atom, suggesting increased charge injection due to stronger excitation. This trend indicates a noticeable nonlinear response of BP to the external field, further suggesting that higher field strengths effectively modulate the doping level and carrier population.

In contrast, the GeSe reveals a similar response shown in Fig. S2. For an excitation field amplitude of $E_0 = 0.005\,V/Å$, the induced photodoping level is approximately 0.07 e/atom, indicating a moderate degree of photoexcited carrier injection. As the field strength increases, the photodoping effect becomes more pronounced. When the field is raised to $E_0 = 0.01\,V/Å$, the density reaches 0.12 e/atom, highlighting the strong dependence of photodoping on the driving field. These findings clearly demonstrate that stronger optical fields can effectively modulate the carrier density, providing a controllable pathway to tune the electronic properties under photoexcitation.

In the absence of photodoping, the pristine BP exhibits lattice constants of 4.43 Å and 3.32 Å along the zigzag and armchair directions, respectively (Fig. 3a), which aligns with experimental observations[31]. Upon optical excitation with $E_0 = 0.01\,V/Å$, it manifests a considerable anisotropic response of [-1%, 2%], which means an 1% in-plane contraction along the armchair direction and a 2% expansion along the zigzag direction (Fig. 3b). An analogous anisotropic photostrictive response is also observed in bulk GeSe (Fig. 3c-d). Under the same excitation field of $E_0 = 0.01\,V/Å$, GeSe exhibits a clear biaxial deformation of [-1%, 3%] characterized by an approximately 1% contraction along the armchair direction and a 3% expansion along the zigzag direction, indicating that the direction-dependent lattice response is a general feature of orthorhombic layered semiconductors rather than a material-specific anomaly.

Experimentally, Wang *et al.* observed lattice strains of approximately [-2.9%, 5.8%] and superelastic behavior in bulk GeSe, which is on the same order of magnitude with our calculations[16]. They further demonstrated that GeSe can sustain an exceptionally large lattice distortion, up to 12% tensile strain before fracture, enabled by a distinctive shuffle-twinning mechanism. Under optical excitation, the lattice deformations are expected to reach a comparable magnitude, rendering our predicted results experimentally accessible in the regime. In contrast, this strong lattice deformation stands in sharp distinction to ferroelectric BiFeO$_3$, which exhibits a much weaker response of only 0.4%[32].

Likewise, in the nonpolar perovskite SrRuO$_3$, photoinduced lattice expansion is limited to ~1.12%[6]. These comparisons highlight that the anisotropic photostriction effect in 2D orthorhombic materials is remarkably large and serves as a foundation for optoelectronic devices[3].

In 2D orthorhombic semiconductors, the photostrictions of BP and GeSe manifest the analogous response, and also generate similar impacts on their electronic structures and carrier-lattice couplings. Upon excitation, charge redistribution preferentially weakens bonding in the zigzag direction, driving a strong expansion, with a smaller contraction along the armchair axis in BP. In contrast, GeSe manifests a larger intrinsic bandgap and stronger dielectric screening, which redistributes photoexcited carriers more evenly among bonding channels. This mechanistic contrast emphasizes how the differences in bonding anisotropy and electronic screening dictate the distribution of lattice strains. Whereas BP presents a noticeable photostrictive response arising from its electronic anisotropy, GeSe offers a platform where both the sign and magnitude of the armchair strain can be programmed via excitation density in optomechanical devices.

The considerable anisotropic photostriction provides a foundation for exploring nonequilibrium strain dynamics, strong light-matter interactions, and photoinduced phase transitions. With the advent of pump-probe techniques and scanning probe microscopy under optical excitation, it is possible to resolve the spatiotemporal evolution of photostriction at nanometer-femtosecond scales[8]. In optoelectronic applications, anisotropic photostriction facilitates the realization of ultrathin, light-controlled actuators for adaptive optics and photonic crystals[3]. It provides a route to build light-driven switches with precise control over actuation direction.

**Anisotropic electronic structure and optical absorption.** To elucidate the impact of biaxial strain, the band structures are calculated for three representative configurations: isotropic compression of -1%, isotropic tension of 2%, and anisotropic strain of [-1%, 2%], as shown in Fig. S2. The pristine BP exhibits a bandgap of 0.15 eV. Isotropic compressive strain reduces the bandgap to 0.01 eV, whereas isotropic tensile strain increases it to 0.40 eV. The anisotropic strain leads to a bandgap of 0.22 eV, underscoring the tunability of electronic structures. For GeSe, the strain of [-1%, 3%] also increases the gap from 0.86 eV to 0.90 eV. These results demonstrate that both isotropic and anisotropic strains contribute to an effective means to modulate the electronic properties.

In addition, we calculate the optical absorption spectra of bulk BP and GeSe. The dielectric function is obtained from the electronic structure using the hybrid functional (HSE06)[33]. Fig. 4 shows the absorption spectra along the armchair and zigzag directions, respectively. In pristine BP, the calculated band edge of 0.25 eV closely matches previous experimental observation[34], confirming the reliability of our calculations. When the strain is applied along the armchair direction, the optical response becomes highly sensitive: the first absorption peak undergoes a substantial blue shift, rising from 0.25 eV to 0.38 eV (Fig. 4a). This shift reveals the strong coupling between lattice deformation and the electronic states responsible for low-energy optical transitions. In comparison, the zigzag-polarized response is quite different. The initial absorption feature appears at a much higher energy, approximately 2.71 eV, and the influence of strain is minimal, producing only a slight blue shift of about 0.06 eV (Fig. 4b). These contrasting behaviors underscore the inherently anisotropic nature of the optical properties.

A similar energy-shifted trend is observed in GeSe. The optical absorption follows the monotonic increase in the bandgap and demonstrates the extensive tunability of optical transitions, particularly along the armchair direction. Under the biaxial strain of [-1%, 3%], the first absorption peak shifts markedly from 1.34 eV to 1.68 eV as the strain increases for the armchair polarization (Fig. 4c), indicating effective modulation of the low-energy excitations. In contrast, for the zigzag polarization, the absorption onset remains at a higher energy near 2.02 eV and exhibits only a modest blue shift of roughly 0.05 eV (Fig. 4d). The comparison between the two materials highlights how directional strain can serve as an efficient and selective mechanism for tuning their anisotropic optical responses, especially in low-symmetry layered semiconductors. In previous GW plus Bethe-Salpeter equation simulations, researchers found that the exciton binding energy of bulk BP is only 0.03 eV[35], revealing a weak exciton effect in the optical properties.

**Photostriction modulated carrier dynamics.** To further investigate the impact of lattice deformation on carrier dynamics, we perform nonadiabatic molecular dynamics simulations[36-38]. By explicitly accounting for the electron-phonon interactions and lattice deformations, these simulations allow real-time tracking of electron-hole recombination and reveal how the biaxial strains determine the balance between radiative and nonradiative process[39, 40].

Time-domain simulations are presented for the pristine structure as well as the photoinduced configurations, where the biaxial strains are [-1%, 2%] for BP and [-1%, 3%] for GeSe, respectively. Fig. 5 displays the evolution of the electron population on CBM and the time scale of recombination. As a benchmark, the recombination timescale of pristine BP aligns well with the previous theory and experiment[41, 42]. The electron-hole recombination lifetime $\tau$ is quantified by fitting the exponential function $exp(-t/\tau)$. For pristine BP, the electron-hole recombination lifetime is calculated to be 49 ps, consistent with previous time-domain simulations (28 ps)[41] and transient-absorption experiments (~100 ps)[42]. When the biaxial strain of [-1%, 2%] is applied, the lifetime increases to 190 ps, representing a fourfold enhancement relative to the unstrained system. For GeSe, the carrier lifetimes exhibit a similar modulation under strains, increasing from 8.29 ns in the pristine material to 16.20 ns at the biaxial strain of [-1%, 3%]. This nearly twofold increase highlights the strong strain sensitivity of the recombination dynamics. The extended lifetimes indicate particularly improved electronic efficiency and longer-lived excited states, which benefit optoelectronic performance in strained semiconductors.

To further explore the mechanism of prolonged nonradiative recombination lifetime, we analyze the electron density of states and nonadiabatic coupling strengths. The anisotropic strain shifts the CBM and causes a rise of the bandgap of 0.22 eV in BP, and increases the bandgap from 0.86 eV to 0.90 eV in GeSe (Fig. S3). This variation suggests associated changes in interatomic spacing and orbital overlap. The same trend of reduced coupling strength is found in both the BP and GeSe. The strength shows a modest but systematic reduction, decreasing from 2.16 to 1.72 meV in BP and from 0.59 to 0.50 meV in GeSe, reflecting a weakened coupling under strain. This reduction underscores the critical role of the lattice modulation in mitigating electron-phonon interactions, thereby reducing nonradiative recombination and enhancing quantum efficiency through promoting more favorable alignment of electronic states and suppressing defect-mediated losses[37]. Therefore, the results find biaxial strain engineering to be a versatile strategy for tuning excitonic dynamics and improving optoelectronic performance in layered semiconductor systems.

In order to uncover the microscopic mechanism underlying the carrier dynamics, Fig. S4 shows that the lattice deformation redistributes charge density in BP and GeSe, diminishing charge overlap and directly modulating their electronic structures. Furthermore, the energy evolution in real-time propagation is analyzed (Fig. S5). In BP, the biaxial strain suppresses fluctuations of the Kohn-Sham orbital energies, with a similar reduction observed for GeSe. Vibration spectra are obtained via Fourier transforms of the band-gap autocorrelation functions, with peak intensities indicating the phonon modes contributing to recombination. The strains suppress fluctuations of the orbital energies, indicating enhanced stability of the electronic states. It also modulates the phonon-mediated recombination pathways, revealing a strong coupling between electronic structures and lattice vibrations.

By incorporating electron-phonon interactions, we find that the carrier-lattice coupling-induced strain prolongs the carrier lifetime through suppressing nonradiative recombination. The increased electron-hole recombination lifetime originates from the strain-reduced nonadiabatic coupling strength, enlarged bandgap, and modulated phonon modes. The results further indicate that the lattice deformation under photoexcitation stabilizes excited states, modulates electron-phonon interactions, and improves quantum efficiency. The mechanism underlies photo-strain engineering as a viable strategy for tuning carrier dynamics and enhancing the optoelectronic performance of layered semiconductors.

The anisotropic photostriction in 2D orthorhombic semiconductors opens a route to develop optomechanical devices with unprecedented functional tunability. The highly direction-dependent lattice deformation under optical excitation enables mechanically reconfigurable elements. This opens a pathway toward ultrathin, remote-controllable actuators and robotic systems operating at the nanoscale. Furthermore, the in-plane anisotropy inherent to orthorhombic structure can be harnessed for polarization-sensitive mechanical responses, allowing for the design of programmable optical elements. The fast, reversible, and non-volatile nature of the photostrictive effect also suggests its integration into neuromorphic computing architectures and memory elements, where the mechanical deformation encodes information or modulates circuit behavior. The properties position the 2D materials as a versatile class of functional media for light-driven actuation, adaptive photonics, and nanoelectromechanical systems.

## Conclusion.

In summary, we demonstrated an anisotropic photostriction effect and strain-modulated carrier dynamics in 2D orthorhombic semiconductors through time-dependent density functional theory, capturing the nonequilibrium carrier-lattice dynamics under photoexcitation. Light-induced lattice deformations in 2D orthorhombic materials not only exhibit large strain but are also strongly direction-dependent, arising from the interplay between electronic redistribution and elastic anisotropic structure. The anisotropic response manifests as an expansion along the armchair direction with a simultaneous contraction along the zigzag direction, directly from the intrinsic structural anisotropy of the layered orthorhombic materials. The photostriction responses are highly tunable with photodoping density, allowing precise control over both the magnitude and orientation. The anisotropic strains enlarge the bandgap and blueshift the absorption spectra, especially along the armchair direction. Substantially, the photo-induced anisotropic strains significantly extend carrier recombination lifetimes in GeSe and BP by suppressing nonradiative recombination, primarily due to the weakened nonadiabatic coupling. These findings provide microscopic insight into anisotropic optomechanical coupling in 2D orthorhombic systems and establish a foundation for designing ultrathin, light-responsive actuators and reconfigurable optoelectronic devices.

## Acknowledgments

This work was supported by the National Natural Science Foundation of China

(12574255). The numerical calculations in this study were partially carried out on the ORISE Supercomputer.

## Author contributions

J.Z. designed the research. All authors contributed to the analysis and discussion of the data and the writing of the manuscript.

## Conflict of Interest

The authors declare no competing financial interest.

## Methods

**Time-dependent density functional theory.** The TDDFT calculations were performed utilizing the time-dependent *ab initio* package (TDAP), developed based on the time-dependent density functional theory[22, 23, 26] and implemented within SIESTA[24, 25, 27, 43]. The dynamic simulations were carried out with an evolving time step of 50 as for both electrons and ions within a micro-canonical ensemble. To explain the TDDFT methods more, we present a more detailed description of our methods based on time-dependent Kohn-Sham equations for coupled electron-ion motion. We perform ab initio molecular dynamics for coupled electron-ion systems with the motion of ions following the Newtonian dynamics, while electrons follow the time-dependent dynamics. The ionic velocities and positions are calculated with the Verlet algorithm at each time step. When the initial conditions are chosen, the electronic subsystem may populate any state, ground or excited, and is coupled nonadiabatically with the motion of ions. We carried out k-space integration using a 10×12×4 mesh for the bulk case in the Brillouin zone of the supercell to confirm the convergence in real-time propagation. The electron density [$\rho^{OCDFT}(r)$] in the general form of an orbital-constrained DFT expression is:

$$\rho^{OCDFT}(r) = \sum_i n_i^{OCDFT} \phi_i^*(r)\phi_i(r) \quad (2)$$

The total energy for the orbital-constrained DFT is written as $E_{KS}[\rho^{OCDFT}(r)]$ instead of ground-state $E_{KS}[\rho_0(r)]$. Herein $E_{KS}[\rho]$ is based on the orbital-constrained electron density, and $\rho_0(r)$ is the electron density in the ground state $\rho_0(r) = \sum_i n_i \phi_i^*(r)\phi_i(r)$; $n_i$ is the occupation number of the *i*-th orbital in ground state and $n_i^{OCDFT}$ is the fixed target occupation for the *i*-th orbital for the excited states. We elucidate the lattice deformation driven by photodoping carriers in layered orthorhombic semiconductors. Structural dynamics under photoinduced carrier occupations are evaluated using the constrained density functional theory with fixed orbital occupations[43-44].

**Real-time carrier dynamics.** The optimized atomistic and electronic structures are calculated by density functional theory (DFT) using the Vienna ab initio simulation package (VASP)[28]. The electronic wave function is described using projector-augmented wave (PAW) pseudopotential with the exchange-correlation interaction treated using Perdew-Burke-Ernzerhof (PBE) method[29, 30, 45]. The energy cutoff for the plane-wave basis was set to 500 eV for all calculations. For the calculations involving BP and GeSe systems, the 160-atom and 145-atom supercells are utilized, respectively. And the Γ point is used for Brillouin-zone sampling. The nonadiabatic molecular dynamics (NAMD) simulation is carried out using Hefei-NAMD, which combines the time-dependent Kohn-Sham equation with the fewest switches surface hopping method[46-49].

**Absorption calculations.** The absorption spectra were calculated from the dielectric function using the expression $\alpha(\omega) = \frac{\sqrt{2}\omega}{c}\left[\sqrt{\varepsilon_1^2 + \varepsilon_2^2} - \varepsilon_1\right]^{\frac{1}{2}}$, $\varepsilon_1$ and $\varepsilon_2$ are the real and imaginary parts of the dielectric function, ω is the light frequency, c is the speed of light in vacuo[50]. The weak van der Waals interactions between different layers are considered by the vdW-DF level with the optB88 exchange functional (optB88-vdW)[51, 52]. Excitonic contributions were not considered in our calculations. The total number of bands considered was set to be twice that used in the total-energy and band structure calculations. Because the dielectric function is a tensor, the absorption spectra along the armchair and zigzag directions were obtained separately.

# Figures and Captions

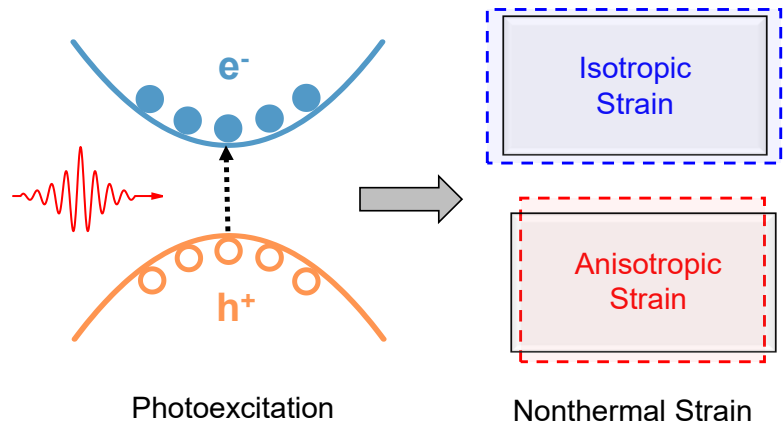

**Figure 1. Schematic illustration of the photostriction effect in semiconductors.** Upon optical excitation, electrons in the valence bands are excited to the conduction bands. The resulting nonthermal strain, arising from electron-lattice coupling, is referred to as photostriction. Both isotropic (upper panel) and anisotropic (lower panel) strain responses are depicted.

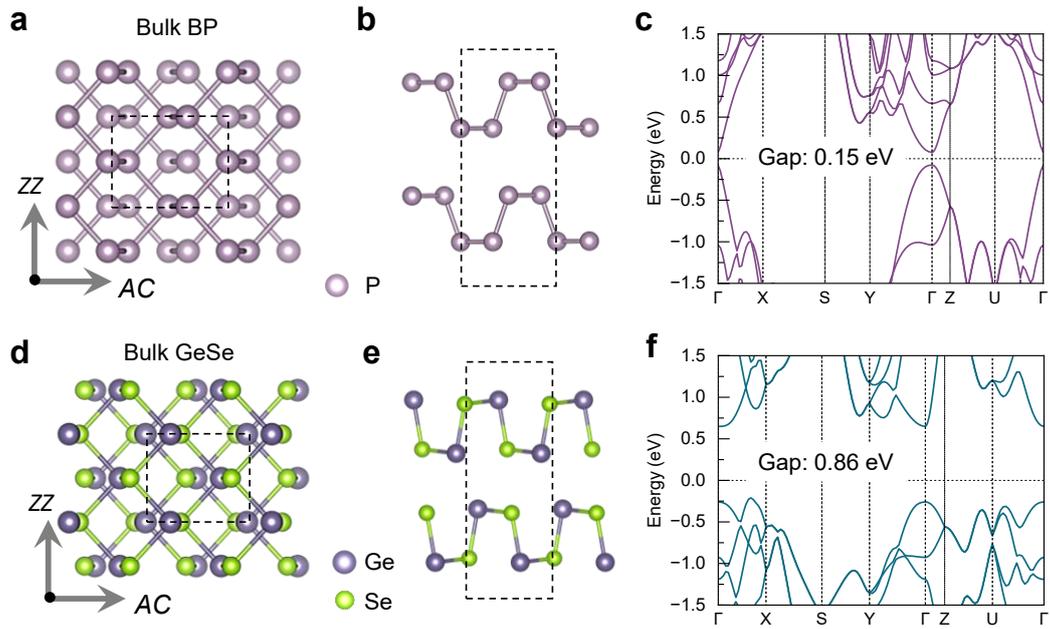

**Figure 2. Geometric and electronic structures of bulk BP and GeSe.** (a) Top and (b) side view of bulk BP, with lattices shown in dashed lines. (c) Band structure of bulk BP, with the Fermi level set to zero. (d) Top and (e) side view of bulk GeSe. (f) Band structure of bulk GeSe, with the Fermi level set to zero. The armchair (*AC*) and zigzag (*ZZ*) directions are indicated in panels (a) and (d). The band structures are calculated using the PBE functional.

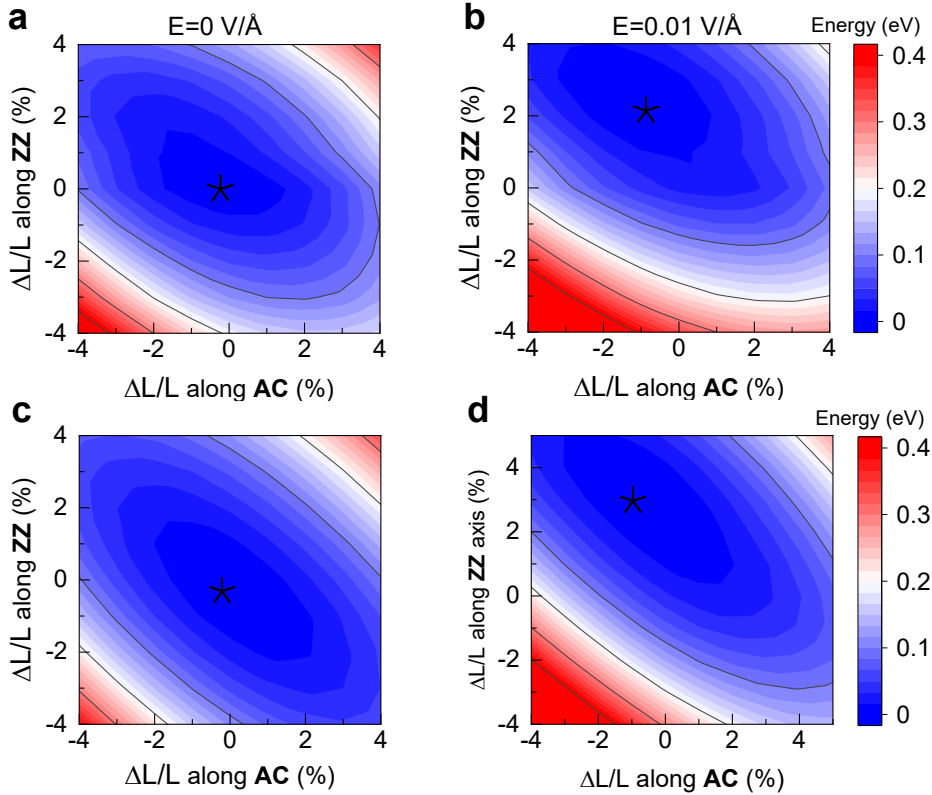

**Figure 3. Photodoping carrier-induced lattice deformation in bulk BP and GeSe.** (a) Energy diagrams of bulk BP without photoexcitation with different lattice deformations. The energy minimum is located at lattice constants of *a* = 4.43 Å and *b* = 3.32 Å along the armchair (AC) and zigzag (ZZ) directions, respectively. (b) The same quantity as (a) under the low electric field strength of 0.01 V/Å. The energy minimum is located at [-1%, 2%], which indicates 1% contraction along the armchair and 2% extension along zigzag directions, respectively. (c) Energy diagrams with different lattice deformations of GeSe without photoexcitation. (d) The same quantity as (c) under a low electric field strength of 0.01 V/Å. The energy minimum is located at [-1%, 3%]. The energies are recalculated relative to the minimum in each panel. The stars denote he energy minima in each panel.

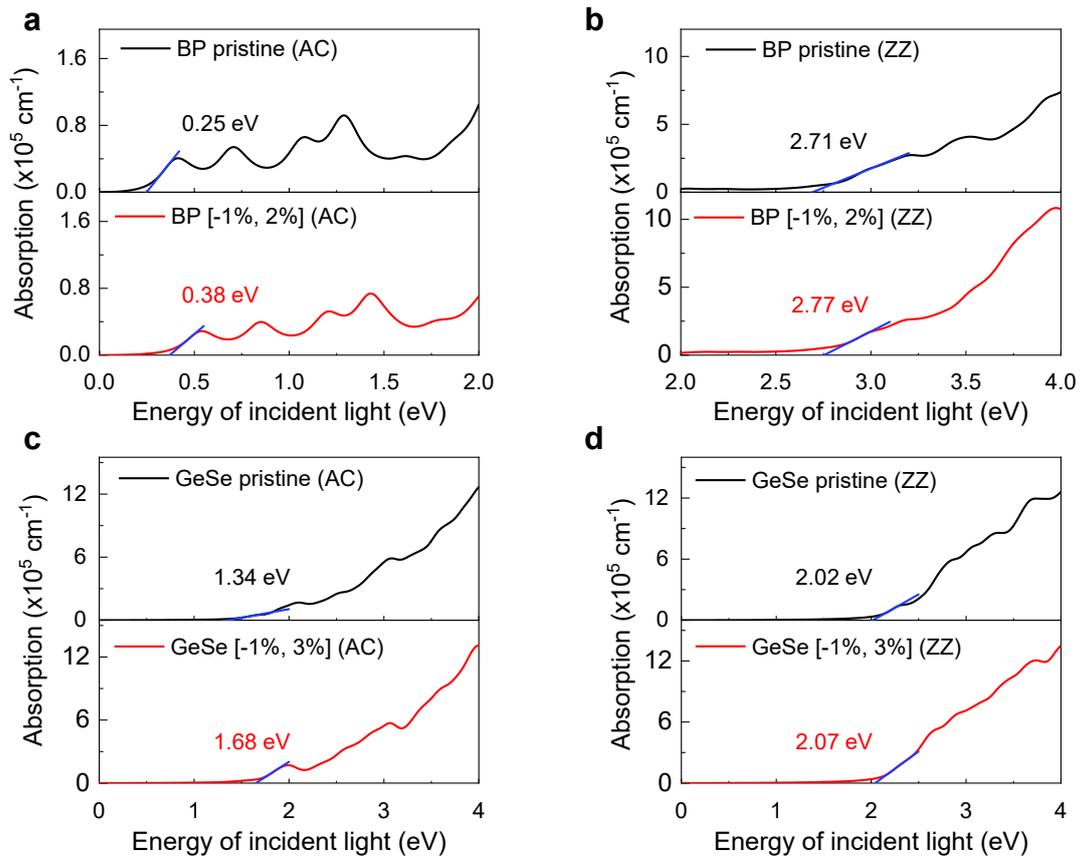

**Figure 4. Photostriction-modulated optical absorption spectra in bulk BP and GeSe.** Optical absorption spectra of bulk BP (a, b) with light incident along the perpendicular axis and polarized along the armchair and zigzag directions, respectively. Optical absorption spectra of bulk GeSe (c, d) with light incident along the perpendicular axis and polarized along the armchair and zigzag directions, respectively. Blue dashed lines show linear fits to estimate the band edges of the first absorption peak, revealing anisotropic light absorption in response to biaxial strains between the armchair and zigzag axes.

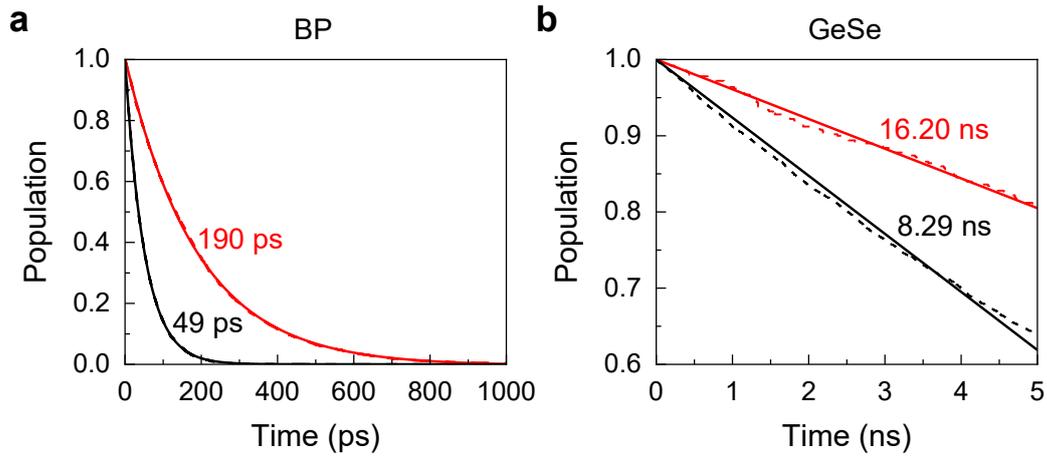

**Figure 5. Strain-modulated suppression of nonradiative recombination in bulk BP and GeSe.** (a) Dynamics of the photoexcited electron population at the conduction band minimum (CBM) for pristine (black) and biaxial strain [-1%, 2%] in BP (red). The carrier lifetime obtained with an exponential fit is prolonged in the strained BP (190 ps) compared to the pristine one (49 ps). (b) Time evolution of the photoexcited electron population at CBM in GeSe for the pristine (black) and biaxially strained [-1%, 3%] (red) cases. The carrier lifetime, extracted from an exponential fit, is substantially extended under the strain (16.20 ns) relative to the pristine material (8.29 ns). The original data are plotted in dashed lines, and the fitted curves are shown in solid lines.

**Table of Contents**

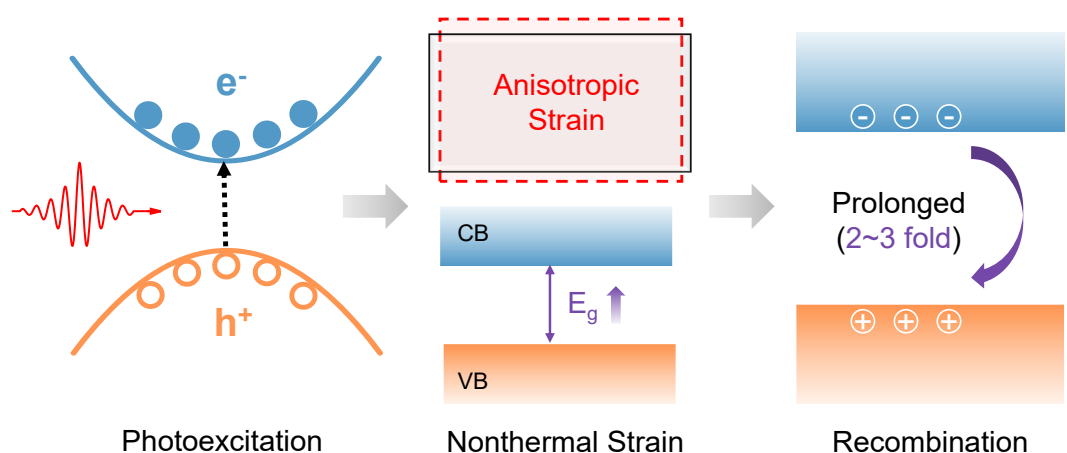

# Supplemental Materials for
# "Anisotropic Photostriction and Strain-modulated Carrier Lifetimes in Orthorhombic Semiconductors"

Jianxin Yu[1,+], Kun Yang[1,+], Jiawen Li[1,+], Sheng Meng[2,3], Xinghua Shi[1], Jin Zhang[1,*]

[1]Laboratory of Theoretical and Computational Nanoscience, National Center for Nanoscience and Technology, Chinese Academy of Sciences. Beijing 100190, P. R. China

[2]Beijing National Laboratory for Condensed Matter Physics, and Institute of Physics, Chinese Academy of Sciences, Beijing 100190, P. R. China

[3]Songshan Lake Materials Laboratory, Dongguan, Guangdong 523808, P. R. China

\* Corresponding author: Jin Zhang (jinzhang@nanoctr.cn)

**This file contains：**

Note. Regarding TDDFT methods.

S1. Laser-induced carriers in BP and GeSe.

S2. Lattice and electronic structure of BP with different strains.

S3. Strain-induced density of states analysis.

S4. Strain-induced charge redistributions in BP and GeSe.

S5. Time-dependent energy evolution and vibrational analysis.

**Note. Regarding TDDFT methods**

We present more description on our methods based on time-dependent Kohn-Sham equations for coupled electron-ion motion[1-6]. We can perform ab initio molecular dynamics for coupled electron-ion systems with the motion of ions following the Newtonian dynamics, while electrons follow the time-dependent Kohn-Sham equation. The ionic velocities and positions are calculated with the Verlet algorithm at each time step. When the initial conditions are chosen, the electronic subsystem may populate any state, ground or excited, and is coupled nonadiabatically with the motion of ions.

This approximation works well for situations where a single path dominates in the reaction dynamics, for the initial stages of excited states before significant surface crossings take place, or for cases where the state-averaged behavior is of interest when many electron levels are involved in condensed phases. However, the approach has limitations when the excited states involve multiple paths, especially when state-specific ionic trajectories are of interest. In such cases, Ehrenfest dynamics fails as it describes the nuclear path by a single average point even when the nuclear wave function has broken up into many different parts. In this work, the deficiency is not critical because the laser-induced path dominates in the whole reaction dynamics. In addition, we have tested the dynamics by altering the initial conditions, such as ionic temperatures and laser intensities, to confirm the robustness of our conclusion.

For infinite periodic systems, we treat the laser field with a vector potential,

$$\vec{A}(t) = -c \int^{t} \vec{E}(t')dt' . \qquad (1)$$

The time evolution of the wavefunctions $\{\psi_i(\mathbf{r}, t)\}$ are then computed by propagating the Kohn-Sham equations in atomic units (a.u.),

$$i\frac{\partial}{\partial t}\psi_i(\mathbf{r}, t) = \left[\frac{1}{2m}\left(\vec{p} - \frac{e}{c}\vec{A}\right)^2 + V(\mathbf{r}, t)\right]\psi_i(\mathbf{r}, t) , \qquad (2)$$

where $V(\mathbf{r}, t)$ is expressed as:

$$V(\mathbf{r}, t) = \sum V_I^{local}(\mathbf{r}) + \sum V_I^{KB} + V^H(\mathbf{r}) + V^{xc}(\mathbf{r}). \qquad (3)$$

Here, $V_I^{local}$ and $V_I^{KB}$ are the local and Kleinman–Bylander parts of the pseduopotential of ion I, and $V^H$, $V^{xc}$ are the Hartree, exchange-correlation (XC) potential, respectively. As the ions are much heavier than electrons by at least three orders of magnitude, the nuclear positions are updated following the Newton's second law,

$$M_I \frac{d^2 \mathbf{R}_I}{dt^2} = \sum_i \left\langle \psi_i \left| \nabla_{\mathbf{R}_I} (\frac{1}{2m}\left(\vec{p} - \frac{e}{c}\vec{A}\right)^2 + V(\mathbf{r},t)) \right| \psi_i \right\rangle, \qquad (4)$$

where $M_I$ and $\mathbf{R}_I$ are the mass and position of the $I^{th}$ ion, respectively. Equation (3) and (5) represents the time-dependent coupled electron-ion motion. The time-dependent Kohn-Sham equations of electrons and the Newtonian motion of ions are solved simultaneously, with ionic forces along the classical trajectory evaluated through the Ehrenfest theorem.

The forces acting on the ions can be calculated through the Hellmann–Feynman theorem[1]:

$$F_J = -\nabla_{R_J} \sum_{K \neq J} \frac{Z_J Z_K}{|R_J - R_K|} - \sum_n f_n \left[ c_n^* \left(\nabla_{R_J} H\right) c_n - \sum_{i,j} c_n^{*i} \left\langle \chi_i \left| \nabla_{R_J}(V_H + V_{xc}) \right| \chi_j \right\rangle c_n^j - c_n^*(HS^{-1}d^J + h.c.)c_n \right], \qquad (5)$$

where $R_J$ and $F_J$ are the position and force of J-th ion. $V_H$ and $V_{xc}$ are the Hartree and exchange-correlation potentials, respectively, and $d^J$ is the vector matrix $d_{ij}^J = \left\langle \chi_i \left| \nabla_{R_J} \right| \chi_j \right\rangle$. In the case of adiabatic dynamics, the last term in brackets reduces to the ordinary Pulay force term, $-f_n \varepsilon_n c_n \left(\nabla_{R_J} S\right) c_n$, with $\varepsilon_n$ being he eigenvalue of the n-th Kohn-Sham wavefunction.

The dependence of the number of excited electrons on laser strength $E_0$ is calculated by projecting the time-evolved wavefunctions ($|\psi_{n,\mathbf{k}}(t)\rangle$) on the basis of the ground-state wavefunctions ($|\varphi_{n',\mathbf{k}}\rangle$),

$$\eta(t) = \frac{1}{n_e}\frac{1}{N_k} \sum_{n,n'}^{CB} \sum_k^{BZ} |\langle \psi_{n,\mathbf{k}}(t)|\varphi_{n',\mathbf{k}}\rangle|^2, \qquad (6)$$

where $N_{\mathbf{k}}$ is the total number of the **k**-points used to sample the Brilluoin zone and $n_e$ is the total number of electrons. The sum over the band indices n and n' run over all conduction bands and the maximum value of $\eta(t)$ is recorded.

    To mimic photoexcitation of different laser pulses, we can also elaborately build the initial electronic states upon photoexcitation. We change the population of Kohn-Sham orbitals from the ground state to specific electronic configurations. The population variation is proportional to the optical transition probability. From the figures below, it can be seen that the occupations and spatial charge distributions upon photoexcitation calculated based on this approach are in good agreement with those using the real-time laser fields.

## S1. Laser-induced carriers in BP and GeSe.

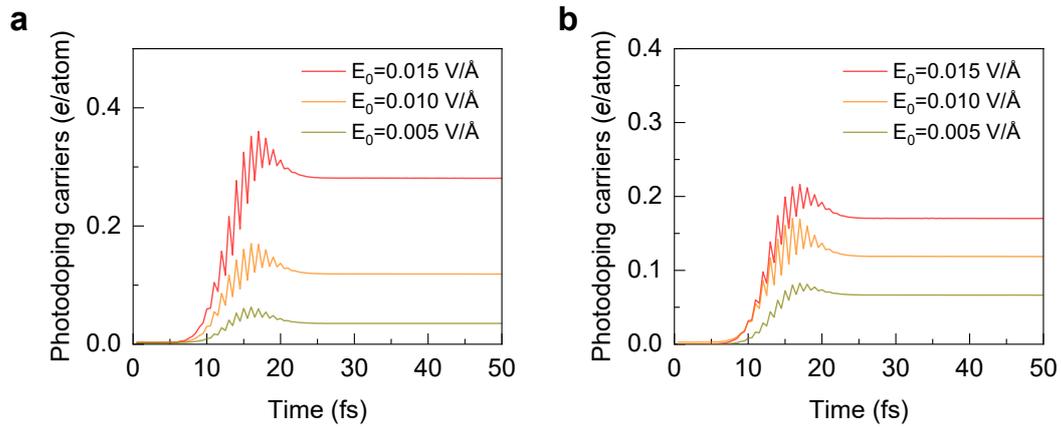

**Figure S1.** The number of excited electrons upon photoexcitation from the valence bands to conduction bands under photoexcitation (a) for BP and (b) for GeSe. The applied electric fields are along the in-plane direction. The photo energy is set to 2.0 eV, and for the laser strength is $E_0$= 0.005, 0.010, and 0.015 V/Å, respectively. The laser field value reaches the maximum strength $E_0$ at time t=15 fs and ends at 30 fs, respectively.

## S2. Lattice and electronic structure of BP with different strains.

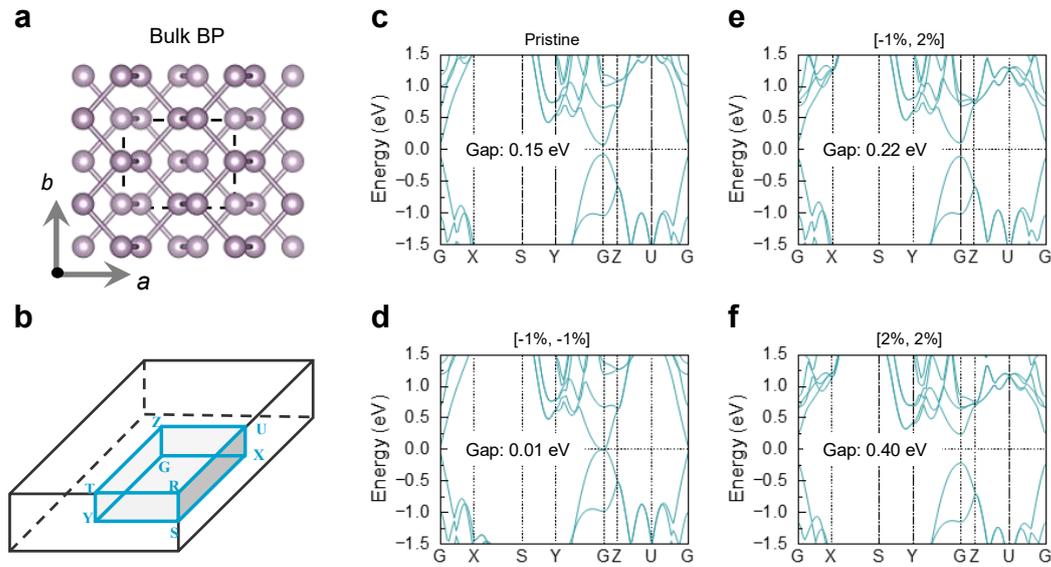

**Figure S2. Lattice and electronic structure of BP with different strains.** (a) Top view of the atomic structure in bulk BP. Dashed lines and (*a*, *b*) mark the lattice and lattice vectors, respectively. (b) The first Brillouin zone in the reciprocal space of the tetragonal unit cell. High-symmetry points and the first Brillouin zone are colored in blue. The tetragonal cell is used in this study in BP. Electronic band structures calculated with Perdew-Burke-Ernzerhof (PBE) functional in (c) pristine, (d) [-1%, 2%], (e) [-1%, -1%], and (f) [2%, 2%]. The Fermi levels are set to 0 eV.

As exhibited in Figure S2, we analyze three biaxial strains: biaxial compression [-1%, -1%], biaxial extension [2%, 2%], and the anisotropic strain [-1%, 2%], respectively. Figure S2(b) shows the reciprocal space of the tetragonal unit cell and high-symmetry points. The bandgap of pristine BP is 0.15 eV. Isotropic biaxial strain [-1%, -1%] results in the smaller bandgap (0.01 eV). In contrast, the bandgap increases to 0.40 eV for the biaxial strain [2%, 2%]. In contrast, anisotropic strain [-1%, 2%] leads to a bandgap of 0.22 eV.

## S3. Strain-induced density of states analysis.

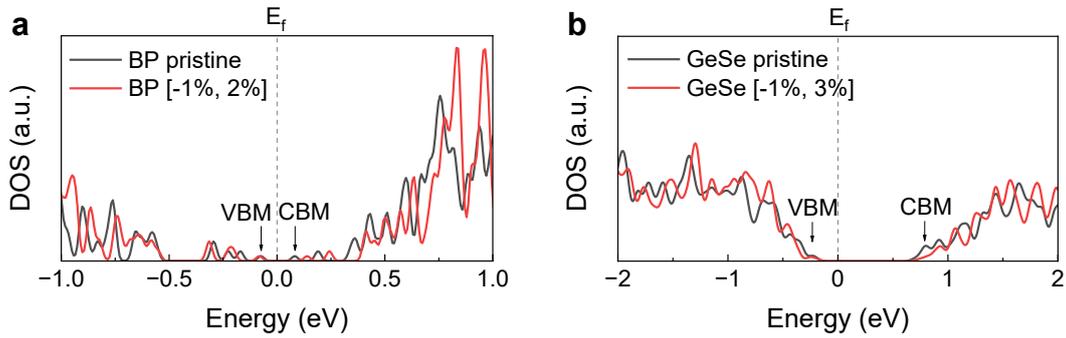

**Figure S3. Strain-induced density of states redistributions of CBM and VBM in BP and GeSe.** The total densities of states between pristine (black) and strained (red) states in (a) BP and (b) GeSe. The pointed CBM and VBM highlight the strain-induced electronic redistribution. The increased bandgaps by strains are mainly contributed to the shift of VBM. The Fermi energies of all structures are set to 0 eV.

## S4. Strain-induced charge redistributions in BP and GeSe.

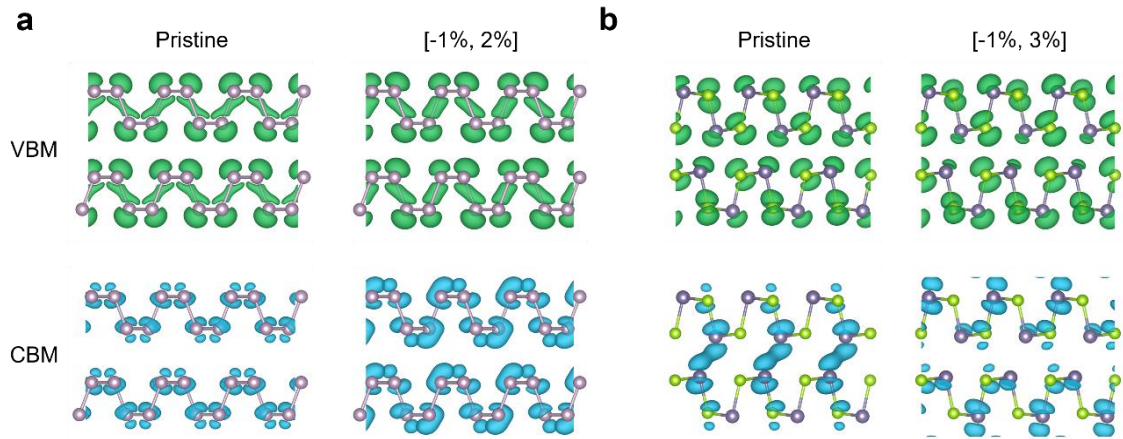

**Figure S4. Strain-induced charge redistributions of CBM and VBM in BP and GeSe.** The charge distribution of CBM and VBM for pristine and strain states in (a) BP and (b) GeSe, respectively. The isosurface values of charge densities are set to $4.3\times10^{-3}$ e/Bohr$^3$.

Figure S4 shows the redistribution of charge density under the anisotropic strains in BP and GeSe. For BP, localized charge distribution around P atoms reflects the large overlap between CBM and VBM in pristine BP. In contrast, the anisotropic strain disperses the charge distribution of CBM and decreases the overlap. Similar redistribution induced by biaxial strains is found in GeSe.

## S5. Time-dependent energy evolution and vibrational analysis.

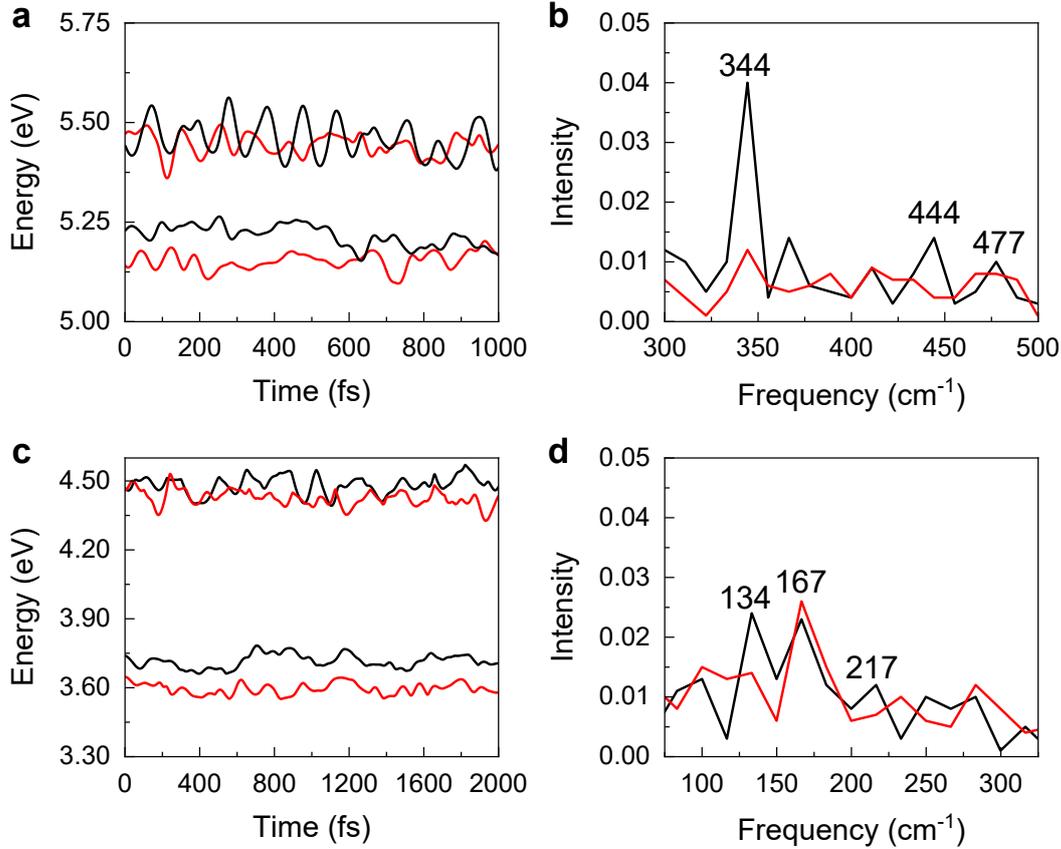

**Figure S5. Time-dependent energy evolution and vibrational analysis in BP and GeSe.** The energy evolution of CBM and VBM between pristine (black) and strained (red) states in (a) BP and (c) GeSe. CBM is located on the higher energy level of every state. The vibrational spectra of pristine (black) and strained (red) states during electron-hole recombination in (b) BP and (d) GeSe. The dominant phonon modes of pristine are marked for clarity.

We analyze the energy evolution of BP and GeSe in the simulations. For BP, the fluctuations of the Kohn-Sham orbital energies are noticeably reduced under anisotropic strain. A similar reduction is observed for GeSe. The vibration spectra are obtained from the Fourier transforms of the band gap autocorrelation functions. The peak intensities indicate the phonon modes involved in recombination. The dominant phonon modes of BP are located at 344 cm$^{-1}$, 444 cm$^{-1}$, and 477 cm$^{-1}$, which are near the experiment results. The Roman modes observed by experiments are out-of-plane vibration of $A^1_g$ (361 cm$^{-1}$), P–P stretching mode $B_{2g}$ (442 cm$^{-1}$) along the zigzag direction, and P–P stretching mode $A^2_g$ (472 cm$^{-1}$) along the armchair direction, respectively.[7] For GeSe, the dominant phonon modes are at 134 cm$^{-1}$, 167 cm$^{-1}$, and 217 cm$^{-1}$, respectively. The peak of 133 cm$^{-1}$ is assigned to the Ge-Se stretching mode of B$_{3g}$ (153 cm$^{-1}$) in the zigzag direction[8]. The 166 cm$^{-1}$ peak corresponds to the out-of-plane vibration mode of A$_g^2$ (179 cm$^{-1}$)[8]. The peak located at 217 cm$^{-1}$ is attributed to the Ge-Se stretching mode of A$_g^3$ (190 cm$^{-1}$) along the armchair direction[8].